\documentclass[sigconf]{acmart}

\usepackage{booktabs}
\usepackage{enumitem}
\usepackage{xspace}
\usepackage{multirow}
\usepackage{microtype}
\usepackage{graphicx}
\usepackage{tabularx}

\newcommand{\eg}{\emph{e.g.,}\xspace}

\newcommand{\etal}{\emph{et al.}\xspace}

\copyrightyear{2026}
\acmYear{2026}
\setcopyright{cc}
\setcctype{by}
\acmConference[EASE '26]{International Conference on Evaluation and Assessment in Software Engineering}{June 09--12, 2026}{Glasgow, United Kingdom}
\acmBooktitle{International Conference on Evaluation and Assessment in Software Engineering (EASE '26), June 09--12, 2026, Glasgow, United Kingdom}
\acmDOI{10.1145/3816483.3830358}
\acmISBN{979-8-4007-2348-3/2026/06}

\begin{document}

\title{From Observation to Explanation: Mechanistic Interpretability of LLM-Generated Code for Principled Repair}

\author{Musfiqur Rahman}
\affiliation{%
  \institution{Concordia University}
  \city{Montr\'eal, Qu\'ebec}
  \country{Canada}}
\email{musfiqur.rahman@mail.concordia.ca}

\renewcommand{\shortauthors}{Rahman}

\begin{abstract}
Large language models (LLMs) have transformed software development, yet the interpretation and understanding of how they generate code remain critically underdeveloped. We tackle this gap by triangulating three layers of evidence.
\emph{Longitudinally}, autonomous agent-attributed code is statistically indistinguishable from human code under a semantic modification definition (HR\,=\,1.019, $p$\,=\,0.490), yet attracts disproportionately corrective modifications in production (49.1\% vs.\ 37.9\% human baseline)-- a finding that demands a causal explanation. \emph{Empirically}, existing benchmarks overestimate LLM capabilities: models that achieve 84--89\% correctness on synthetic tasks attain only 24--35\% on real-world class-level code. \emph{Behaviourally}, detection systems reveal puzzling asymmetries: GPT-3.5 is far more detectable than modern models (AUC-ROC\,=\,0.96, a 27-point
gap), even though modern models produce mutually similar code structures; yet no single feature reliably distinguishes AI-generated from human code at the class level, and the leading commercial detector achieves only 15--64\% recall. Together, these findings raise a question that empirical studies cannot answer: \emph{why} do LLMs generate code with these properties? This thesis proposes \textbf{mechanistic interpretability}--causal token-level circuit tracing--as the bridge between empirical observation and principled correction. By identifying the transformer circuits responsible for detectable stylistic signatures, hallucination patterns, and systematic defects, we aim to drive an agentic repair framework that intervenes at the causal source of defects rather than patching surface symptoms.
\end{abstract}


\maketitle

\noindent\textit{Advisor:}~Emad Shihab
(\href{mailto:emad.shihab@concordia.ca}{emad.shihab@concordia.ca}),
Department of Computer Science and Software Engineering,
Concordia University, Montr\'eal, Qu\'ebec, Canada.

\section{Intorduction}

\subsection{Motivation}

The integration of LLMs into software development has produced a growing body of empirical work that consistently reveals observations without explanations. Our completed studies span three interconnected threads: how agent-attributed code behaves in production, how LLMs perform on real-world generation tasks, and how their outputs can be detected. Each thread surfaces phenomena that the next cannot explain. We present them in the order that motivates the thesis: from the field, to the laboratory, to the black box. The central contribution of this thesis is not any one of these threads in isolation, but the mechanistic account that explains why all three converge on the same unresolved phenomena, and the repair framework that acts on that account.

\subsubsection{Characterizing the Stability of Agent-Attributed Code in Open Source Software}

If LLMs routinely introduce structural defects into production codebases, we would expect agent-attributed code to be unstable--merged and quickly replaced, conforming to the ``disposable code'' narrative that has gained traction in industry reports~\cite{gitclear2023}. Testing this hypothesis requires longitudinal, unit-level evidence that aggregate churn metrics cannot provide.

We applied Kaplan-Meier estimation and Cox Proportional Hazards regression~\cite{rahman2026will}--which estimates each tier's modification risk relative to a human baseline as a hazard ratio (HR)--to 223,997 line-level and 61,729 hunk-level code units across 156 open-source repositories from the AIDev dataset~\cite{li2025rise}, stratified by human-AI collaboration mode:
pair-programmer (Copilot, Cursor, Claude~Code), batch-generator (OpenAI Codex), and an autonomous agent (Devin). Two modification definitions were applied: a syntactic baseline (any content change) and a semantic AST-structural definition using {\tt tree-sitter}. Developer avoidance was tested directly by classifying 10,958 agent-attributed files as co-owned or agent-orphaned via the GitHub Commits API.

\begin{table}[h]
\small
\caption{Code stability by collaboration tier. Mod.\,\%\,=\,share of units
modified at least once during the observation window; Median\,=\,median days to first modification; HR\,=\,Cox hazard ratio for modification risk
relative to the human baseline (HR\,$<$\,1.0 indicates lower risk).
Def\,1\,=\,syntactic modification definition; Def\,2\,=\,AST-structural modification definition.}
\label{tab:stability}
\begin{tabular}{@{}lcccc@{}}
\toprule
\textbf{Tier} & \textbf{Mod.\,\%} & \textbf{Median} & \textbf{HR (Def1)} & \textbf{HR (Def2)} \\
\midrule
Pair-programmer & 51.3\% & 208\,d & 0.959*** & 0.837*** \\
Batch-generator & 51.2\% & 206\,d & 0.790*** & 0.735*** \\
Autonomous      & 68.7\% & 155\,d & 0.899*** & 1.019 \\
\midrule
Human (baseline) & 71.0\% & 101\,d & 1.000 & 1.000 \\
\bottomrule
\multicolumn{5}{@{}l@{}}{\scriptsize ***$p<0.001$;\;
avoidance hypothesis rejected ($p<0.001$)}
\end{tabular}
\end{table}

Table~\ref{tab:stability} presents the central finding. Pair-programmer and batch-generator attributed code is substantially more stable than the human baseline under both definitions. The disposable-code narrative is false for these tiers.

The autonomous agent tier, however, tells a different story. It shows apparent syntactic stability (HR\,=\,0.899) yet is statistically indistinguishable from human code at the AST-structural level
(HR\,=\,1.019, log-rank $p$\,=\,0.490), and it attracts disproportionately corrective modifications (49.1\% vs.\ 37.9\% human baseline, +11.2\,pp). Developer avoidance was empirically ruled out: agent-orphaned files are modified \emph{more} frequently than co-owned files (61.3\% vs.\ 56.7\%, $p<0.001$), the opposite of what avoidance predicts. The simplest interpretation: autonomous agents produce syntactically plausible code that passes initial review but harbours structural defects that surface as post-merge corrections.

This finding directly motivates the thesis. \emph{Something in the generation process of autonomous agents produces structurally defective code.} Empirical observation can characterize the defects; it cannot identify their cause. That causal question requires mechanistic interpretability.

\subsubsection{Beyond Synthetic Benchmarks: Evaluating LLM Performance on Real-World Class-Level Python Code Generation}

Before asking why LLMs produce defective code, we must first establish \emph{how} severely real-world performance diverges from what published benchmarks suggest. We introduced \textbf{RealClassEval}~\cite{rahman2025beyond}, a benchmark of 400 Python classes from real open-source repositories, partitioned into \emph{seen} (repositories pre-dating model training cutoffs) and \emph{unseen} (post-cutoff repositories) sets to test whether memorization explains any performance advantage.

Across seven state-of-the-art LLMs, results reveal a dramatic collapse: models achieving 84--89\% correctness on the synthetic ClassEval benchmark~\cite{du2024evaluating} attain only 24--35\% on
real-world classes. Crucially, performance on seen and unseen repositories is statistically indistinguishable ($p>0.05$), ruling out memorization as an explanation. The gap is structural, not encyclopedic. Documentation provides only a marginal benefit (1--3\%, not statistically significant); RAG is helpful only when specifications are incomplete (4--7\%). Error analysis across all seven LLMs reveals that \texttt{AttributeError}, \texttt{TypeError}, and \texttt{AssertionError} account for 84\% of all failures--the same attribute hallucination pattern visible in the stability study's corrective modification signal.

\subsubsection{OpenClassGen: A Large-Scale Corpus of Real-World Python Classes for LLM Research~\cite{rahman2025openclassgen}}

RealClassEval establishes the gap; studying it at scale requires infrastructure. We released \textbf{OpenClassGen}: 324,843 real-world Python classes extracted from 2,970 engineered open-source projects, each paired with its skeleton and enriched with 27 static code metrics.

\begin{table}[h]
\small
\caption{OpenClassGen evaluation: three LLMs on 300 stratified classes.
CB-F3\,=\,CodeBERTScore-F3; TSED\,=\,tree-edit similarity~\cite{rahman2025openclassgen}.}
\label{tab:openclassgen}
\begin{tabular}{@{}lcccc@{}}
\toprule
\textbf{Model} & \textbf{ROUGE-L} & \textbf{CB-F3} & \textbf{TSED} & \textbf{Pass} \\
\midrule
GPT-o4-mini      & 0.60 & 0.88 & 0.82 & 0.29 \\
Claude-4-Sonnet  & 0.68 & 0.90 & 0.84 & 0.36 \\
Qwen-3-Coder     & 0.67 & 0.90 & 0.84 & 0.35 \\
\midrule
\textbf{Average} & 0.65 & 0.89 & 0.83 & 0.33 \\
\bottomrule
\end{tabular}
\end{table}

Table~\ref{tab:openclassgen} highlights the semantic-functional gap: LLMs achieve strong semantic similarity (CodeBERTScore-F3\,=\,0.89) yet only 0.33 functional correctness. LLMs understand class \emph{intent} but cannot reliably produce \emph{executable} implementations. The dataset supports fine-tuning, difficulty modelling, RAG corpus construction, and failure mode analysis. It is publicly available at~\cite{musfiqur_rahman_2025}.

\subsubsection{Automatic Detection of LLM-Generated Code: A Comparative Case Study Across Function and Class Granularities}

Given that LLM-generated code is structurally defective and performs poorly in real-world settings, reliable detection is essential for academic integrity, intellectual property protection, and quality assurance. We conducted a comparative detection study spanning 14,485 Python functions and 11,913 Python classes generated by four LLMs (GPT-3.5, GPT-OSS,
Claude~3~Haiku, Claude~Haiku~4.5) from CodeSearchNet. Using interpretable structural metrics (18 function-level, 39 class-level) selected via AutoSpearman, we trained CatBoost (CB) classifiers for each model-granularity configuration and benchmarked against GPTZero.

\begin{table}[h]
\small
\caption{Detection AUC-ROC: CatBoost vs.\ GPTZero. GPTZero recall shown in parentheses; CatBoost recall ranges from 61--87\%~\cite{rahman2025detection}.}
\label{tab:detection}
\setlength{\tabcolsep}{4pt}
\begin{tabular}{@{}lcccc@{}}
\toprule
& \multicolumn{2}{c}{\textbf{Function-level}} &
  \multicolumn{2}{c}{\textbf{Class-level}} \\
\cmidrule(lr){2-3}\cmidrule(lr){4-5}
\textbf{Model} & \textbf{CB} & \textbf{GPTZero} &
                 \textbf{CB} & \textbf{GPTZero} \\
\midrule
GPT-3.5          & 0.96 & 0.70\,(29\%) & 0.89 & 0.53\,(15\%) \\
GPT-OSS          & 0.80 & 0.86\,(64\%) & 0.81 & 0.83\,(51\%) \\
Claude 3 Haiku   & 0.70 & 0.71\,(35\%) & 0.83 & 0.62\,(20\%) \\
Claude Haiku 4.5 & 0.71 & 0.76\,(42\%) & 0.80 & 0.66\,(21\%) \\
\bottomrule
\end{tabular}
\end{table}

Table~\ref{tab:detection} reveals three non-obvious findings. (1)~\textbf{The GPT-3.5 anomaly}: GPT-3.5 is 27 percentage points more detectable than any modern model at the function level (AUC\,=\,0.96), a gap that cannot be explained by known structural features alone.
(2)~\textbf{Granularity dominates}: Feature sets overlap at Jaccard\,=\,0.69--0.855 across models at the same granularity, yet collapse to Jaccard\,=\,0.099 across granularities--an 8.6$\times$ drop statistically indistinguishable from random chance, indicating that function-level and class-level detection rely on fundamentally disjoint structural signatures. (3)~\textbf{Commercial recall failure}: GPTZero achieves only 15--64\% recall; our structural CatBoost achieves 61--87\%.
The \emph{Comment-to-Code Ratio} is the sole universal discriminator across all 8 configurations (average SHAP\,=\,0.777), yet zero universal features exist at the class level, and even this feature exhibits opposing directions across model families: older models generate sparser documentation than humans; modern models generate denser documentation.

These findings reveal structural signatures without explaining them. Why is GPT-3.5 anomalously detectable? Why does granularity dominate model architecture in determining which features discriminate? Why is Comment-to-Code Ratio the only universal signal? These are the questions that motivate the interpretability pillar.

\subsubsection{Software Ingestion \& Extraction for Verifiable Evaluation (SIEVE)}

A recurring challenge across all three threads is \emph{data contamination}: if a benchmark, a detection dataset, or an interpretability corpus contains code that an LLM has already seen during training, any downstream finding is suspect. Resolving this empirically is impossible--LLM training corpora are opaque and unverifiable--but it can be mitigated operationally.
We built \textbf{SIEVE} as general-purpose infrastructure for sourcing code from GitHub and automatically classifying it as AI-authored or Human-authored, enabling researchers to construct contamination-aware datasets for any downstream task.
 
SIEVE is backed by a CodeBERT-based binary classifier trained on 114K balanced samples spanning nine LLMs and three programming languages (Python, Java, JavaScript), achieving F1\,=\,0.9478 and
AUROC\,=\,0.9902. The pipeline covers automated GitHub repository discovery, multi-language
code extraction via Tree-sitter AST parsing, cross-repository deduplication, and LLM scoring--all exposed through both a command-line interface and a live web application (\url{sieve-tool.streamlit.app}). The tool has 273 automated tests with 80\% code coverage.
 
The intended use cases of SIEVE extend well beyond detection research. A \emph{generation researcher} can use SIEVE to harvest confirmed LLM-generated functions and classes for benchmarking, ensuring that evaluation targets were not memorized by the model under test.
A \emph{detection researcher} can use it to construct balanced, multi-LLM training corpora without manual curation. A \emph{software evolution researcher} can use it to identify agent-attributed code units in OSS repositories for longitudinal analysis, as in our stability study. A \emph{mechanistic interpretability researcher} can use it to source generation instances paired with known provenance for circuit tracing experiments. In short, SIEVE is not a detection tool that happens to produce data; it is a \emph{data sourcing infrastructure} that happens to use a detection classifier as its core mechanism, and it is what makes the other threads in this thesis contamination-aware rather than merely contamination-conscious.

\subsection{The Unexplained Phenomena}

Read together, the three threads above describe a single arc: production
code behaves unpredictably in ways aggregate metrics miss (stability),
laboratory benchmarks reveal that the underlying models are far less
reliable than published numbers suggest (evaluation), and the resulting
code carries a detectable signature we cannot yet explain (detection).
Each thread produces a solid empirical fact; none of them, individually
or together, explains \emph{why} the phenomenon occurs.
That causal gap is what the remaining pillars of this thesis address.
Four concrete, tractable phenomena anchor this gap.

\begin{enumerate}[leftmargin=*,topsep=2pt,itemsep=2pt]

\item \textbf{The GPT-3.5 anomaly.}
Why does GPT-3.5 produce functions that are 27 percentage points more
detectable than those of modern models on identical prompts?
Which generation circuits produce the extreme brevity and documentation
sparsity that define its structural signature, and why have modern models
evolved away from them?

\item \textbf{The AttributeError paradox.}
LLMs trained on billions of lines of Python never produce SyntaxErrors,
yet they routinely produce AttributeErrors--references to non-existent
object attributes--which dominate 84\% of real-world failures alongside
TypeError and AssertionError.
Syntactic knowledge is encoded in a stable, generalizable circuit.
Object attribute knowledge is not.
What is structurally different about these circuits?

\item \textbf{The token decision problem.}
When generating a function that computes a random value, how does a model
decide to generate \texttt{import numpy} rather than \texttt{import random}?
This is not a question about the surface output; it is a question about
\emph{which internal circuits activate at the moment of token selection}
and why they favour one valid continuation over another.
Preliminary experiments confirm that library aliasing decisions are mediated
by context-dependent retrieval circuits, not knowledge gaps--validating
the tractability of this question.

\item \textbf{The autonomous agent quality gap.}
Autonomous agents produce code that passes syntactic review yet is corrected
at nearly twice the rate of pair-programmer code.
At which token positions in the generation sequence do these structural
defects originate, and which circuits are active at those positions?
\end{enumerate}

\subsection{Thesis Statement}

\emph{The empirical properties of LLM-generated code--its detectability,
its characteristic defect patterns, and its post-deployment fragility--
are causally traceable to specific token-level generation decisions within
transformer circuits.
Identifying these circuits is a necessary prerequisite for principled,
targeted repair of LLM-generated code defects.}

\subsection{Contributions}

\begin{enumerate}[leftmargin=*,topsep=2pt,itemsep=1pt]

\item \textbf{Characterizing the Stability of Agent-Attributed Code in
Open Source Software}~\cite{rahman2025stability}: Survival analysis of 223,997 code units across 156 repositories
disproving the disposable-code narrative while revealing the autonomous
agent structural quality gap.

\item \textbf{Beyond Synthetic Benchmarks}~\cite{rahman2025beyond}: 400-class real-world benchmark establishing a 50--65\,pp performance
collapse across seven LLMs.

\item \textbf{OpenClassGen}~\cite{rahman2025openclassgen}: 324,843-class corpus with 27 static metrics from 2,970 projects. 

\item \textbf{Automatic Detection of LLM-Generated Code}~\cite{rahman2025detection}: First multi-model, multi-granularity detection study with interpretable
features and adversarial robustness evaluation.

\item \textbf{SIEVE}:
Contamination-free data sourcing and detection infrastructure, publicly available as a web interface.

\item \textbf{Mechanistic Interpretability Study}:
Token-level circuit analysis of the four unexplained phenomena above.

\item \textbf{Interpretability-Informed Agentic Repair}:
Repair framework intervening at the causal source of generation defects.
\end{enumerate}

\section{Related Work}

\paragraph{Code Generation Benchmarking.}
HumanEval~\cite{chen2021evaluating} established Pass@k on 164 function-level
tasks representing fewer than 30\% of real-world
programming~\cite{yu2024codereval}.
ClassEval~\cite{du2024evaluating} pioneered class-level evaluation on 100
manually crafted tasks; GPT-4 drops from 85.4\% to 37.0\% Pass@1.
Our work using the RealClassEval dataset extends this to real-world classes
with a seen/unseen contamination split, revealing a further collapse.
HumanEvo~\cite{zheng2025humanevo} addresses temporal contamination at the
function level through repository rollback; OpenClassGen~\cite{rahman2025openclassgen}
complements this at the class-level scale without requiring rollback infrastructure.

\paragraph{LLM-Generated Code Detection.}
Suh \etal~\cite{suh2024empirical} reported the first cross-model study,
achieving up to 82\% F1 within-dataset but only 43--47\% across datasets.
CodeGPTSensor~\cite{xu2025distinguishing} achieved 99.2\% accuracy via
contrastive learning, but on ChatGPT-generated Python only.
Pordanesh \etal~\cite{pordanesh2025hiding} demonstrated that trivial perturbations substantially degrade detection.
Our study is the first systematic multi-model,
multi-granularity comparison with interpretable features, establishing the
8.6$\times$ granularity dominance effect.

\paragraph{Defect Patterns and Code Quality.}
Tambon \etal~\cite{tambon2024bugs} established a taxonomy of ten LLM-specific
bug patterns across four categories (functionality, language, contextual,
hallucination).
Misinterpretation and missing corner cases dominate at 23--24\% and 18--23\%
respectively.
Our findings corroborate and extend this taxonomy: the 84\% failure share
of \texttt{AttributeError}, \texttt{TypeError}, and \texttt{AssertionError} in RealClassEval aligns with the hallucination and contextual categories.
However, the taxonomy is descriptive; our mechanistic interpretability work provides the causal account.

\paragraph{AI Code in the Wild.}
GitClear~\cite{gitclear2023} documented rising aggregate code churn since
2021 coinciding with AI adoption.
Ehsani \etal~\cite{ehsani2026} found that autonomous agent PRs fail CI/CD
and face higher rejection rates.
Our stability study is the first to track
individual code units longitudinally via survival analysis, apply a
semantic AST-structural modification definition, and empirically test the
avoidance alternative hypothesis.

\paragraph{Mechanistic Interpretability.}
Elhage \etal~\cite{elhage2021mathematical} formalised the mathematical
framework of transformer circuits, showing that attention heads implement
operations such as copying and in-context lookup.
Meng \etal~\cite{meng2022locating} localised factual associations in GPT
to specific MLP layers via causal activation patching, enabling targeted
model editing.
Olsson \etal~\cite{olsson2022induction} identified induction heads as
the circuit underlying in-context learning across all transformer models.
In software engineering, circuit-level mechanistic analysis of code LLMs
is virtually absent.
Wan \etal~\cite{wan2022they} applied attention analysis to pre-trained
code models, but stopped at correlation rather than causal intervention.
To the best of our knowledge, this thesis is the first in SE to apply activation patching to identify
causal generation circuits in code LLMs.

\paragraph{Automated Program Repair.}
RepairAgent~\cite{bouzenia2024repairagent} achieved 164 correct fixes on
835 Defects4J bugs via agentic test-driven patch generation.
VulRepair~\cite{fu2022vulrepair} and SelfAPR~\cite{ye2022selfapr} similarly
depend on pre-existing test suites.
All treat the model as a black box and generate repair candidates by
sampling--with no connection between the repair strategy and the
generation decision that caused the defect.
Our repair pillar provides this missing causal link through mechanistic
circuit knowledge.

\section{Research Questions and Challenges}

\paragraph{Pillar~1 -- Evaluation.}
\textbf{RQ1}: How does real-world class-level performance compare to
synthetic benchmarks, and does training-data exposure explain the gap?
\textbf{RQ2}: How does contextual richness across a five-level hierarchy
affect performance on contamination-free generation targets?

\paragraph{Pillar~2 -- Detection.}
\textbf{RQ3}: How do detection accuracy and feature importance vary across
LLMs and code granularities?
\textbf{RQ4}: How robust are structural detectors to adversarial
perturbations (post-hoc comment stripping; same-project few-shot
regeneration)?

\paragraph{Pillar~3 -- Mechanistic Interpretability.}
\textbf{RQ5}: Which circuits drive the detectable stylistic signatures
of GPT-3.5 compared to modern models?
\textbf{RQ6}: What circuit-level differences explain why LLMs preserve
Python syntax yet hallucinate object attributes?
\textbf{RQ7}: How do models resolve token-level ambiguity during library
and API selection, and do these circuits generalize across model families?
\textbf{RQ8}: At which token positions do structural defects characteristic
of autonomous agents originate?

\paragraph{Pillar~4 -- Agentic Repair (Planned).}
\textbf{RQ9}: Can mechanistic knowledge of error-inducing circuits inform
agentic interventions that reduce targeted defect rates without sacrificing
functional correctness?

\paragraph{Challenges.}
Mechanistic findings on small open-weight models may not transfer to
large proprietary models; we mitigate by triangulating across families and
scaling to 7B$+$ variants.
Circuit instability under quantization is addressed by replication across
seeds and context variations.
Repair evaluation beyond Pass@k requires root-cause alignment scoring.
Training contamination is acknowledged as a writing-only limitation,
mitigated by temporal filtering and composite-key deduplication.

\section{Work Plan}

\subsection{Research Approach}

\paragraph{Phase 1 -- Evaluation.}
RealClassEval was constructed by mining GitHub repositories using commit
timestamps to enforce the seen/unseen split.
Functional correctness used automated test suites (Pynguin: 58.7\% branch
coverage on RealClassEval vs.\ 93.2\% on ClassEval--illustrating the
complexity gap).
OpenClassGen applied engineering-quality filters~\cite{munaiah2017curating}
to 2,970 repositories, producing skeletons via AST-based extraction and
enriching each class with 27 Understand static metrics.
RealEval Phase~4 extends evaluation to contamination-free function-level
targets: 653 functions across 50 post-cutoff repositories at five contexts
levels (L0: docstring; L1: stub; L2: prior sibling functions; L3:
dependency-filtered schema; L4: dependency-aware BM25 RAG, $k$\,=\,3).
Generation experiments run on the Narval HPC cluster
(5 LLMs $\times$ 5 context levels, A100-40GB GPUs, temperature\,=\,0,
seed\,=\,42).

\paragraph{Phase 2 -- Detection.}
The comparative study used a global intersection methodology: only snippets generated successfully by all four LLMs were retained, ensuring
differences reflect model characteristics rather than task difficulty.
CatBoost was selected via Scott-Knott ESD test across 30 repetitions of
10-fold cross-validation (300 measurements per configuration).
An adversarial robustness experiment is underway:
\emph{Condition~A} tests post-hoc comment stripping
(inline only / docstrings only / both, three-way ablation);
\emph{Condition~B} tests the same-project three-example few-shot regeneration
($\pm$50\% LOC constraint, 3-tier prompt fallback, $n$\,=\,385 per cell).
Pickled CatBoost models are reused without retraining.

\paragraph{Phase 3 -- Mechanistic Interpretability.}
Mechanistic interpretability asks not \emph{what} a model outputs but
\emph{which internal components causally determine that output}.
We apply four complementary techniques via TransformerLens:

\emph{Logit lens}: Project the residual stream at each transformer layer
onto the vocabulary and record token probability distributions.
This reveals whether a hallucinated attribute name crystallizes early in
the network (implicating embedding or early attention layers) or late
(implicating MLP factual recall layers).

\emph{Activation patching}: Run generation twice--clean prompt and
corrupted version (\eg replacing an object context with an alternative)--
and selectively substitute activations between runs at specific
positions (attention heads, MLP layers, residual stream).
The causal effect of each substitution identifies which components are
necessary and sufficient for the observed token decision.

\emph{Attention head attribution}: Gradient-weighted rollout across
attention heads identifies which heads mediate specific token decisions.
For library selection (\eg \texttt{import numpy} vs.\ \texttt{import random}),
we expect early-layer contextual retrieval heads to be responsible.

\emph{Linear probing}: Train lightweight classifiers on hidden state
activations at each layer to predict defect type (Tambon \etal\ taxonomy
as ground truth labels).
High probe accuracy at a specific layer localizes the representation of
defect-inducing semantics.

A preliminary finding validates this programme:
in Qwen2.5-Coder-1.5B and DeepSeek-Coder-1.3B,
library aliasing decisions (\eg \texttt{cv2} vs.\ \texttt{opencv})
are mediated by \emph{context-dependent retrieval circuits},
not knowledge gaps--contradicting the intuition that aliasing errors
reflect missing training data.
Activation patching confirmed that substituting the context shifts the model's token prediction even when both aliases name the same library.

Planned experiments will: (i)~scale to 7B-class open-weight code models
(\eg Qwen2.5-Coder-7B, DeepSeek-Coder-6.7B, CodeLlama-7B);
(ii)~connect circuit findings directly to RQ5--RQ8 by targeting the token
positions corresponding to anomalous detectability, attribute
hallucinations, and library selection;
(iii)~compare circuits across these model families to identify universal
versus model-specific patterns.

\paragraph{Phase 4 -- Agentic Repair (Planned).}
This pillar is the thesis's terminal deliverable: it operationalizes
Phase~3's circuit-level findings into targeted interventions, and is
evaluated not only on whether defects decrease but on whether the
intervention acts on the same circuit Phase~3 identifies as causal.
The framework will operate in three steps:
(1)~run Phase~3 circuit analysis during defect generation to identify
the responsible circuit and token position;
(2)~select an intervention type based on defect category--activation
steering (directly modifying the residual stream at the responsible
circuit) for hallucination defects, a mechanistically-guided prompt
reformulation for context retrieval failures;
(3)~evaluate the repair against Pass@k and a root-cause alignment score
that verifies whether the targeted circuit matches the identified
error-inducing circuit.

This is qualitatively different from RepairAgent~\cite{bouzenia2024repairagent}
and SelfAPR~\cite{ye2022selfapr}, which treats the model as a black box
and generate patch candidates by sampling.
Our framework treats generation as a white box and applies
causally-motivated interventions--requiring no test suite and producing
repairs that are traceable to a mechanistic diagnosis.

\subsection{Metrics (GQM)}
Table~\ref{tab:gqm} shows the metrics we use to quantify the results of our various experiments.

\begin{table}[h]
\small
\caption{Goal-Question-Metric alignment.}
\label{tab:gqm}
\begin{tabular}{@{}p{1.5cm}p{2.5cm}p{3.2cm}@{}}
\toprule
\textbf{Goal} & \textbf{Question} & \textbf{Metric} \\
\midrule
Evaluate real-world generation &
How does context richness affect contamination-free performance? &
Pass@k, functional coverage, edit distance \\
\addlinespace
Assess detection robustness &
How does adversarial perturbation affect structural detectors? &
F1, AUROC, recall under Conditions A and B \\
\addlinespace
Explain generation mechanisms &
Which circuits cause each defect pattern? &
Logit attribution, patching causal effect (Cohen's~$d$) \\
\addlinespace
Validate mechanistic repair &
Does circuit-informed intervention reduce defects? &
Pass@k delta, defect reduction rate, root-cause alignment \\
\bottomrule
\end{tabular}
\end{table}

\subsection{Validity Threats and Mitigation}

We organize threats following the standard construct, internal, external,
and conclusion validity framework~\cite{wohlin2012experimentation}.
The most fundamental threat is benchmark contamination: LLM training
corpora are opaque and unverifiable, so we cannot empirically confirm
that the evaluation targets are unseen.
We mitigate it operationally through temporal filtering and SIEVE-based
sourcing rather than claiming it is resolved.
For the interpretability pillar, a distinct threat arises from studying
small open-weight models: circuit findings at 1.5B--7B parameters may not
transfer to the large proprietary models (GPT-4.1, Claude~4) that
produced the empirical anomalies motivating RQ5--RQ8.
We address this by triangulating across two model families and scaling
experiments to 7B$+$ variants.
The remaining threats and their mitigations are summarised in
Table~\ref{tab:threats}.

\begin{table}[h]
\small
\caption{Validity threats and mitigations.}
\label{tab:threats}
\begin{tabular}{@{}p{1.8cm}p{0.9cm}p{3.4cm}@{}}
\toprule
\textbf{Threat} & \textbf{Type} & \textbf{Mitigation} \\
\midrule
Benchmark contamination &
Internal &
Temporal filtering; composite-key dedup; writing-only limitation acknowledged \\
Model non-determinism &
Internal &
Temperature\,=\,0; fixed seed (42); reproducibility verified \\
Small-model generalization &
External &
Scale to 7B$+$; triangulate across two model families \\
Circuit instability &
Construct &
Replicate across seeds, prompts, and context variants \\
Repair metric adequacy &
Construct &
Pass@k $+$ root-cause alignment $+$ human evaluation \\
LLM evolution &
External &
Modular pipeline; temporal bounds explicitly acknowledged \\
\bottomrule
\end{tabular}
\end{table}

\subsection{Progress and Trajectory}
 
The first two pillars of this thesis, evaluation and detection, are
substantially complete; the third, mechanistic interpretability, is
underway; and the fourth, agentic repair, is planned for the final
year.
 
The in-the-wild stability study is complete, and its longitudinal
survival analysis is being prepared for wider dissemination.
The evaluation pillar's real-world class-level benchmark is complete and
under review, supported by a large-scale corpus that is already published.
The detection pillar's multi-model, multi-granularity study is likewise
complete and under review, and the contamination-free data-sourcing
infrastructure that underlies all three empirical threads is built and
publicly available.
 
Two efforts are underway now: an adversarial robustness experiment
testing the detection classifiers against realistic evasion attempts,
and the mechanistic interpretability work, which is scaling beyond the
small models used in the pilot study while targeting the four unexplained
phenomena above (RQ5--RQ8).
 
Looking ahead, the thesis will extend the detection pillar with a
contrastive-learning study, complete the mechanistic interpretability
pillar across model families, and close with the design, implementation,
and evaluation of the interpretability-informed agentic repair
framework---turning circuit-level explanation into targeted correction.

\begin{acks}
We used AI-powered tools (Claude and Grammarly) to assist with revisions of this submission. All AI-assisted content was reviewed and verified for accuracy by the author.
\end{acks}

\bibliographystyle{ACM-Reference-Format}
\bibliography{references}

@inproceedings{chen2021evaluating,
  author    = {Mark Chen and others},
  title     = {Evaluating Large Language Models Trained on Code},
  year      = {2021},
  note      = {arXiv:2107.03374}
}

@inproceedings{du2024evaluating,
  author    = {Xueying Du and others},
  title     = {Evaluating Large Language Models in Class-Level Code Generation},
  booktitle = {Proc. ICSE},
  pages     = {1--13},
  year      = {2024}
}

@article{rahman2025beyond,
  author    = {Musfiqur Rahman and SayedHassan Khatoonabadi and Emad Shihab},
  title     = {Beyond Synthetic Benchmarks: Evaluating {LLM} Performance on
               Real-World Class-Level {Python} Code Generation},
  journal   = {Empirical Software Engineering},
  year      = {2025},
  note      = {Under revision}
}

@inproceedings{rahman2025openclassgen,
  author    = {Musfiqur Rahman and SayedHassan Khatoonabadi and Emad Shihab},
  title     = {{OpenClassGen}: A Large-Scale Corpus of Real-World {Python}
               Classes for {LLM} Research},
  booktitle = {Proc. EASE},
  year      = {2026},
  doi       = {10.1145/3816483.3816547}
}

@article{rahman2026will,
  title={Will It Survive? Deciphering the Fate of AI-Generated Code in Open Source},
  author={Rahman, Musfiqur and Shihab, Emad},
  journal={arXiv preprint arXiv:2601.16809},
  year={2026}
}

@book{wohlin2012experimentation,
  title={Experimentation in software engineering},
  author={Wohlin, Claes and Runeson, Per and H{\"o}st, Martin and Ohlsson, Magnus C and Regnell, Bj{\"o}rn and Wessl{\'e}n, Anders and others},
  volume={236},
  year={2012},
  publisher={Springer}
}

@article{munaiah2017curating,
  title={Curating github for engineered software projects},
  author={Munaiah, Nuthan and Kroh, Steven and Cabrey, Craig and Nagappan, Meiyappan},
  journal={Empirical Software Engineering},
  volume={22},
  number={6},
  pages={3219--3253},
  year={2017},
  publisher={Springer}
}

@misc{musfiqur_rahman_2025,
	author       = { Musfiqur Rahman and SayedHassan Khatoonabadi and Emad Shihab },
	title        = { OpenClassGen (Revision 0f572f3) },
	year         = 2025,
	url          = { https://huggingface.co/datasets/mrahman2025/OpenClassGen },
	doi          = { 10.57967/hf/6321 },
	publisher    = { Hugging Face }
}

@article{rahman2025detection,
  author    = {Musfiqur Rahman and SayedHassan Khatoonabadi and
               Ahmad Abdellatif and Emad Shihab},
  title     = {Automatic Detection of {LLM}-Generated Code: A Comparative
               Case Study of Contemporary Models Across Function and Class
               Granularities},
  journal   = {Empirical Software Engineering},
  year      = {2025},
  note      = {Under major revision}
}

@article{rahman2025stability,
  author    = {Musfiqur Rahman and Emad Shihab},
  title     = {Characterizing the Stability of Agent-Attributed Code in
               Open Source Software},
  journal   = {ACM Trans. Softw. Eng. Methodol.},
  year      = {2026},
  note      = {In preparation}
}

@inproceedings{yu2024codereval,
  author    = {Hao Yu and others},
  title     = {{CoderEval}: A Benchmark of Pragmatic Code Generation with
               Generative Pre-Trained Models},
  booktitle = {Proc. ICSE},
  pages     = {1--12},
  year      = {2024}
}

@misc{suh2024empirical,
  author    = {Hyunjae Suh and others},
  title     = {An Empirical Study on Automatically Detecting {AI}-Generated
               Source Code: How Far Are We?},
  year      = {2024},
  note      = {arXiv:2411.04299}
}

@article{xu2025distinguishing,
  author    = {Xiaodan Xu and others},
  title     = {Distinguishing {LLM}-Generated from Human-Written Code by
               Contrastive Learning},
  journal   = {ACM Trans. Softw. Eng. Methodol.},
  volume    = {34},
  number    = {4},
  pages     = {1--31},
  year      = {2025}
}

@misc{pordanesh2025hiding,
  author    = {Saman Pordanesh and others},
  title     = {Hiding in Plain Sight: On the Robustness of {AI}-Generated
               Code Detection},
  booktitle = {Proc. DIMVA},
  pages     = {44--64},
  year      = {2025}
}

@misc{tambon2024bugs,
  author    = {F. Tambon and others},
  title     = {Bugs in Large Language Models Generated Code: An Empirical
               Study},
  year      = {2024},
  note      = {arXiv:2403.08937}
}

@misc{gitclear2023,
  author    = {{GitClear}},
  title     = {Coding on {Copilot}: 2023 Data Suggests Downward Pressure
               on Code Quality},
  year      = {2023},
  howpublished = {\url{https://www.gitclear.com/coding_on_copilot_data_shows_ais_downward_pressure_on_code_quality}}
}

@misc{ehsani2026,
  author    = {R. Ehsani and others},
  title     = {Where Do {AI} Coding Agents Fail? An Empirical Study of
               Failed Agentic Pull Requests in {GitHub}},
  year      = {2026}
}

@article{elhage2021mathematical,
  author    = {Nelson Elhage and others},
  title     = {A Mathematical Framework for Transformer Circuits},
  journal   = {Transformer Circuits Thread},
  year      = {2021}
}

@inproceedings{meng2022locating,
  author    = {Kevin Meng and others},
  title     = {Locating and Editing Factual Associations in {GPT}},
  booktitle = {Proc. NeurIPS},
  year      = {2022}
}

@misc{olsson2022induction,
  author    = {Catherine Olsson and others},
  title     = {In-Context Learning and Induction Heads},
  year      = {2022},
  note      = {Transformer Circuits Thread}
}

@inproceedings{wan2022they,
  author    = {Yao Wan and others},
  title     = {What Do They Capture? A Structural Analysis of Pre-Trained
               Language Models for Source Code},
  booktitle = {Proc. ICSE},
  pages     = {2377--2388},
  year      = {2022}
}

@inproceedings{zheng2025humanevo,
  title={Humanevo: An evolution-aware benchmark for more realistic evaluation of repository-level code generation},
  author={Zheng, Dewu and Wang, Yanlin and Shi, Ensheng and Zhang, Ruikai and Ma, Yuchi and Zhang, Hongyu and Zheng, Zibin},
  booktitle={2025 IEEE/ACM 47th International Conference on Software Engineering (ICSE)},
  pages={1372--1384},
  year={2025},
  organization={IEEE}
}

@article{li2025rise,
  title={The rise of ai teammates in software engineering (se) 3.0: How autonomous coding agents are reshaping software engineering},
  author={Li, Hao and Zhang, Haoxiang and Hassan, Ahmed E},
  journal={arXiv preprint arXiv:2507.15003},
  year={2025}
}

@misc{bouzenia2024repairagent,
  author    = {Islem Bouzenia and Premkumar Devanbu and Michael Pradel},
  title     = {{RepairAgent}: An Autonomous, {LLM}-Based Agent for
               Program Repair},
  year      = {2024},
  note      = {arXiv:2403.17134}
}

@inproceedings{fu2022vulrepair,
  author    = {Michael Fu and others},
  title     = {{VulRepair}: A {T5}-Based Automated Software Vulnerability
               Repair},
  booktitle = {Proc. ESEC/FSE},
  pages     = {935--947},
  year      = {2022}
}

@inproceedings{ye2022selfapr,
  author    = {He Ye and others},
  title     = {{SelfAPR}: Self-supervised Program Repair with Test
               Execution Diagnostics},
  booktitle = {Proc. ASE},
  pages     = {1--13},
  year      = {2022}
}

\end{document}